\documentclass[10pt, conference, compsocconf]{IEEEtran}
\usepackage{multirow}
\usepackage{graphics}
\usepackage{pifont}
\usepackage{subfig}
\usepackage{footmisc}
\usepackage{threeparttable}

\usepackage{cite}

\ifCLASSINFOpdf
 \usepackage[pdftex]{graphicx}
\else
\fi
\begin{document}

\newcommand{\tick}{\ding{52}}
\newcommand{\cross}{\hspace{1pt}\ding{55}}
%
\title{V-BOINC: The Virtualization of BOINC}



%
\author{\IEEEauthorblockN{Gary A. McGilvary\IEEEauthorrefmark{1},
Adam Barker\IEEEauthorrefmark{2},
Ashley Lloyd\IEEEauthorrefmark{3} and 
Malcolm Atkinson\IEEEauthorrefmark{1}}
\IEEEauthorblockA{\IEEEauthorrefmark{1}Edinburgh Data-Intensive Research Group, 
School of Informatics, The University of Edinburgh 
\\ Email: gary.mcgilvary@ed.ac.uk, mpa@staffmail.ed.ac.uk}
\IEEEauthorblockA{\IEEEauthorrefmark{2}School of Computer Science, University of St Andrews\\
Email: adam.barker@st-andrews.ac.uk}
\IEEEauthorblockA{\IEEEauthorrefmark{3}Business School, The University of Edinburgh\\
Email: ashley@edinburgh.ac.uk}}


\maketitle

\begin{abstract}
The Berkeley Open Infrastructure for Network Computing (BOINC) is an open source client-server middleware system created to allow projects with large computational requirements, usually set in the scientific domain, to utilize a technically unlimited number of volunteer machines distributed over large physical distances. However various problems exist deploying applications over these heterogeneous machines using BOINC: applications must be ported to each machine architecture type, the project server must be trusted to supply authentic applications,  applications that do not regularly checkpoint may lose execution progress upon volunteer machine termination and applications that have dependencies may find it difficult to run under BOINC.

To solve such problems we introduce virtual BOINC, or V-BOINC, where virtual machines are used to run computations on volunteer machines. Application developers can then compile their applications on a single architecture, checkpointing issues are solved through virtualization API's and many security concerns are addressed via the virtual machine's sandbox environment. In this paper we focus on outlining a unique approach on how virtualization can be introduced into BOINC and demonstrate that V-BOINC offers acceptable computational performance when compared to regular BOINC. Finally we show that applications with dependencies can easily run under V-BOINC in turn increasing the computational potential volunteer computing offers to the general public and project developers.
\end{abstract}


%
\IEEEpeerreviewmaketitle

\section{Introduction}
Volunteer computing, made popular by BOINC and SETI@Home \cite{Anderson2004}  gives members of the general public the opportunity to offer their computational resources to distributed scientific research projects. Created in 2002, BOINC has become the most popular volunteer computing middleware system, where 900,000 users actively participate in projects in the areas of medicine, physics, biology and many more \cite{Anderson2010}. Despite its popularity, BOINC still has many drawbacks, most of which relate to BOINC applications running in the user space of the volunteer machine; the portion of system memory where user processes execute. These drawbacks are as follows: 

\begin{itemize}
\item Project developers are required to port their application to every target machine architecture.
\item Project developers need to provide application-level checkpointing to ensure job progress is not lost upon host termination or failures.
\item Project developers are limited to creating applications that have no dependencies.
\item Users of BOINC must trust that project servers they attach to, will not distribute malicious or untrustworthy applications.

\end{itemize}
These drawbacks can result in project developers taking additional time to implement measures to solve such problems. 

With virtualization, many of these issues are solved. One only needs to port an application to a single virtual machine architecture, host security in which the host is protected from third party applications is inherently addressed by the sandbox environment and system-level checkpointing is available. Applications with dependencies can also easily run where dependencies may be pre-installed or attached to a virtual machine enabling application developers to deploy more complex applications to obtain results of more value. In this paper we present virtual BOINC, or V-BOINC that introduces virtualization into the BOINC framework. Many users within the volunteer community have taken advantage of V-BOINC and information on how to do so, can be found at \cite{vboinc_info}.

The foundation of our approach relies on sending lightweight virtual machine images to volunteer clients allowing BOINC applications to run within the virtual machine itself rather than in the user space of the host. This is implemented by installing a BOINC client within the virtual machine image to fetch applications for a user specified project. This is in addition to the BOINC client installed on the user's host to download the virtual machine image. Our approach to virtualization within BOINC allows V-BOINC to run typical BOINC projects such as SETI@Home and future projects with applications that have dependencies. This will in turn increase the number of potential applications volunteer infrastructures are able to execute. The use of V-BOINC hence aims to enable access to computations that could not otherwise be performed enabling more science, design and business to be done. 

The rest of this paper is organized as follows: next we give an overview of related research and then give a brief overview of BOINC and outline the implementation of V-BOINC in Section 3. In Section 4, we offer our evaluation of V-BOINC specifically comparing the performance of V-BOINC to BOINC and determining the effectiveness of our checkpointing approach. We conclude with a summary of our approach and results in Section 5.

\section{Related Work}
Several other important research projects have added virtualization to BOINC. This section reviews this research while paying specific attention to the differences between our own approach and others.

Ferreira \textit{et al} \cite{Ferreira2011} aim to provide solutions to BOINC's downfalls --- namely porting applications to all participant machines and security --- by employing a virtualization approach to create a BOINC middleware component, for use with VMWare and VirtualBox, called \textit{libboincexec}. Their implementation shows the virtualization approach increases the execution time of an application by 196 seconds for VMWare \cite{vmware_player} and 229 seconds for VirtualBox \cite{vbox} on average, when compared to running the same application via the BOINC framework. 

While the authors achieve good results, their implementation assumes a virtual machine image is already present and is configured correctly on the volunteer machine and no application dependencies exist. The authors show that in order to run a job within the virtual machine, the application must first be transferred to the host machine and copied to the virtual machine. Similarly, output data must be transferred to the BOINC server via the host machine. This method may however introduce security weaknesses where an application and data can be corrupted before they are copied to the virtual machine and vice versa. Furthermore, when an application and its data are large in size, transferring these to the host and then the virtual machine will also further increase the job pre-execution time significantly. 

The authors implementation also breaks the BOINC policy of being transparent to the user where many changes are required to the host due to the external dependencies of \textit{libboincexec}. Also the effects of virtual machine checkpointing, for example the time to create a snapshot and the storage requirements on a volunteer host are not explored; we cover these items in the following sections.



González \textit{et al} \cite{Gonzalez2008} realize that running interpreted applications in BOINC (e.g R, Matlab, Java etc) is difficult when firstly, an application has lots of dependencies and secondly, it is not possible to send an application environment such as Matlab to a host. Currently the BOINC Wrapper exists allowing legacy applications to be run within BOINC, however the authors go further and create a \textit{starter} tool that detects whether the correct environment is present for the application to run successfully and if not, detects missing parts and downloads them. The environment is then deleted after the computation has finished. One problem may however occur if URLs of packages change overtime. The authors also realize that interpreted applications do not have application-level checkpointing and hence introduce virtualization via VMware Player to provide system-level checkpointing. By using VMware Player, users of the authors system will be presented with the virtual machine, violating the BOINC policy of being transparent to the user. In our case, VirtualBox is used allowing headless virtual machines; virtual machines that do not display a window at runtime. Furthermore, the authors use virtual machine checkpointing however its effects on a volunteer host are not explored.

Recently, BOINC offered virtual machine functionality \cite{vboxwrapper} via its \textit{vboxwrapper} program that acts as an interface between the BOINC client and VirtualBox. This program as well as the application and its data are stored in a shared folder between the host and guest, where the computation is then executed. Our approach differs as virtual machine images can be automatically downloaded to the host and execute applications from any BOINC project. The authors method may be useful for typical scientific BOINC applications with no dependencies, however our approach also targets applications with dependencies where we also try to customize and open up BOINC such that researchers and organizations can make use of V-BOINC easily and effectively. 

Similarly, developers at LHC CERN have developed the CernVM that runs data analyses from LHC experiments \cite{Buncic2008}. The virtual machine image is available to run on many hypervisors such as VirtualBox, KVM, VMware, Xen and Hyper-V Server. The CernVM/VBoxWrapper Test Project \cite{cernvm_vbox} is similar to our project where virtual machine images can be downloaded to execute computations, however the framework is not customizable to the point where users are able to select the project they would like to join; only LHC computations can be performed. Their server implementation is also not available where V-BOINC's is publicly available and the V-BOINC virtual machine image size is smaller than the CernVM reducing the transfer time between server and client.

\begin{table*}[ht]
\begin{center}
\caption{Virtualization Technology vs V-BOINC Requirements}
\begin{threeparttable}
\begin{tabular}{|l|c|c|c|}
	\hline
$\textbf{Requirement}$ &  $\textbf{QEMU/KVM}$  & $\textbf{VirtualBox}$ & $\textbf{VMWare Player}$ \\ \hline
Unique IP Address Allocation & \tick\tnote{1} & \tick &  \tick  \\ \hline
Headless VM & \tick &  \tick & \tick\tnote{1} \\ \hline
Image Size $<$ 235MB (compressed) & \tick  &\tick & \tick  \\ \hline
Boot Time $<$ 20s & \tick\tnote{2}  & \tick & \tick  \\ \hline
Basic VM Control & \tick & \tick & \tick\tnote{1} \\ \hline
Remote Command Execution & \cross & \tick & \tick\tnote{1}   \\ \hline
Checkpointing & \tick  & \tick & \tick \\ \hline
Portability (Mac \& Linux) & \tick\tnote{3}  &  \tick & \cross   \\ \hline
\end{tabular}
\begin{tablenotes}
       \item[1] additional configuration and/or installation required on host
       \item[2] only when used with KVM enabled
	\item[3] KVM component not available on Mac OS X
     \end{tablenotes}
\end{threeparttable}
\end{center}
\end{table*}

\section{Virtualizing BOINC}
V-BOINC is the virtualized version of BOINC allowing users to avoid the drawbacks of BOINC and take advantage of virtualization. Implementing the framework requires some additions to convert regular BOINC into V-BOINC. Namely the V-BOINC project server distributes virtual machine images as opposed to scientific applications and the V-BOINC client controls not only the host's BOINC core client but the virtual machine and it's inner BOINC client. These components are relatively difficult to create and hence they can be downloaded alongside their source code on the V-BOINC page at \cite{vboinc_info}; this paper discusses the latter implementation.

A typical BOINC infrastructure has two major components: the project server and the BOINC client installed on the volunteer machine. The BOINC client is composed of 4 parts: the core client, the \textit{boinccmd} interface, the BOINC Manager and the Screensaver. The BOINC core client is the most important component which communicates with the server to deal with user registration, attaches clients to projects and sets up the computation. In order to control the BOINC core client via the command line, the \textit{boinccmd} interface can be used to issue commands such as obtaining new tasks, suspending computations, uploading results etc.

For most BOINC users, many will interact with the BOINC Manager which gives a GUI representation of the command line version, giving the user an easy method to take control of the BOINC client. Similar to the \textit{boinccmd} interface, the BOINC Manager can also set user preferences such as storage and network restrictions --- many other options exist but for brevity, are not explained here. Finally, the Screensaver component offers a project specific screensaver displaying graphics for a running task however whether a screensaver exists is project dependent. 

When the BOINC client is downloaded, the user must select or enter a project to attach to. These projects reside on `BOINC servers' that have the BOINC server environment and dependencies installed; the server uses MySQL for data storage, while Apache and PHP are used for web access (e.g access to a user's online account). The BOINC server has the tasks of distributing, collecting and storing completed jobs from many clients. Upon a user attaching to a project, the server will handle the user registration and record what machine type the user has in order to supply it with the correct executable for its architecture. The computation is setup according to user based preferences, the application runs and the results are sent to the server which are then validated and stored. Despite the basic appearance of the server component, the core concepts behind this are much more complex where several daemons execute and cooperate with one another to provide a reliable and scalable service. These however are out of scope of this paper. We now know enough about the basic operation of BOINC to introduce virtualization into the framework.

 \subsection{Virtualization Technologies}
Firstly we must define the characteristics we require of the virtualization software package. These requirements are listed in Table 1 alongside the three most relevant virtualization technologies and whether they satisfy our conditions. The software packages chosen are QEMU/KVM, VirtualBox and VMWare Player. Other technologies were analysed and were either deemed to be unfit for our purposes or did not provide enough functionality. 

We require that these virtualization technologies allow bridged networking to give the virtual machine a unique IP address enabling the virtual machine's inner BOINC client to directly receive jobs and return results to the BOINC project server. The chosen package must also adhere to the  BOINC policy of being transparent to the user, offer API's for basic virtual machine control (e.g. start, stop, etc) and allow command execution on the virtual machine.

Furthermore virtual machine checkpointing must be available and the chosen package must be portable to both Linux and Mac OS X machines; the platforms V-BOINC targets. Future work will include Windows platforms. Finally, we specify that the virtual machine image must boot within a small period of time, currently under 20 seconds, and that the size of the virtual machine image file while compressed is less than 235 MB which is the current size of the CernVM --- the project most similar to ours.

\begin{itemize}
\item \textit{QEMU/KVM:} QEMU is an open source virtual machine emulator that achieves reasonable performance \cite{Bartholomew2006} \cite{Siever2005}. This performance can be increased by using the Kernel Virtual Machine (KVM) component recently merged into QEMU, that takes advantage of hardware-assisted virtualization via the extensions Intel VT-X or AMD-V \cite{Fisher-Ogden2006} found on recent Linux kernels. QEMU/KVM satisfies the majority of our requirements however it does not offer an API for executing commands upon the guest. 

Furthermore, to obtain a unique IP address, QEMU/KVM requires configuration changes and additional installations on the host which are unreasonable to ask a volunteer user to undertake. The resulting virtual machine does satisfy our boot time requirement in 11 seconds however only when the KVM component is enabled to increase performance; this component is not available on Mac OS X. Without the use of KVM, the performance of the virtual machine would decrease significantly. 

\item \textit{VirtualBox:} is an x86 and AMD64/Intel64 open source virtualization product developed and maintained by Oracle that can be run on all major platforms and supports many guest operating systems \cite{vbox}. VirtualBox does however have components based on QEMU  \cite{Marosi2010} hence it inherently satisfies the same requirements, however in this case, VirtualBox boots up the same image on the same host in approximately 13 seconds. Most importantly, the major advantage of VirtualBox is the ability to easily start the virtual machine image with a Network Bridge Adapter via Ethernet or wireless giving the machine a unique IP address and identity. The VirtualBox API called VBoxManage, also simplifies the task of controlling the virtual machine where QEMU's equivalent provides less relevant options and remote commands can be executed upon the guest via the \textit{guestcontrol} function.

\item \textit{VMWare Player:} is a free virtualization package developed and maintained by VMWare, however this is not open source like the previous packages \cite{vmware_player}. VMWare Player satisfies all requirements apart from being available on both Mac OSX and Linux however allowing headless virtual machines, basic control and remote command execution depend on whether the VIX API is installed.
\end{itemize}

Based on the evidence shown here, with ease of use in mind and to avoid additional installation of packages and configuration on the volunteer host, the most suitable candidate for use within V-BOINC is VirtualBox; V-BOINC currently supports VirtualBox version 4.1.8 however later versions should also work but remain untested. In the future, V-BOINC will support the above hypervisors to increase the user base of this volunteer computing paradigm. We now give an overview of how V-BOINC operates and runs computational jobs.

\subsection{Methodology Overview}
The foundation of V-BOINC relies upon five components each shown in Figure 1:
\begin{itemize}
\item \textbf{V-BOINC Server:} A modified BOINC server distributing virtual machine images, as opposed to scientific applications, to attached volunteer hosts.
\item \textbf{V-BOINC Client:} A downloadable package encapsulating a modified BOINC client and a GUI with the purpose of communicating with the V-BOINC and BOINC Servers as well as the host virtualization hypervisor. 
\item \textbf{The Virtual Machine (VM):} The platform the BOINC scientific application will execute upon. The V-BOINC virtual machine uses the Ubuntu Server 11.04 Operating System (OS) and runs upon a VirtualBox Virtual Disk Image (VDI). A single OS is currently used for initial deployment of the project to the volunteer user community however we envisage an extensive variety in the future. By default, the V-BOINC virtual machine is set to use at most 1 GB of RAM and 1 processor.
\item \textbf{BOINC Server:} A typical BOINC project server that provides scientific applications to attached volunteer hosts.
\item \textbf{Dependency Disks (DepDisk):} A separate VDI containing the application's dependencies.
\end{itemize}

\begin{figure}[h!]
  \begin{center}
\includegraphics[width=0.51\textwidth]{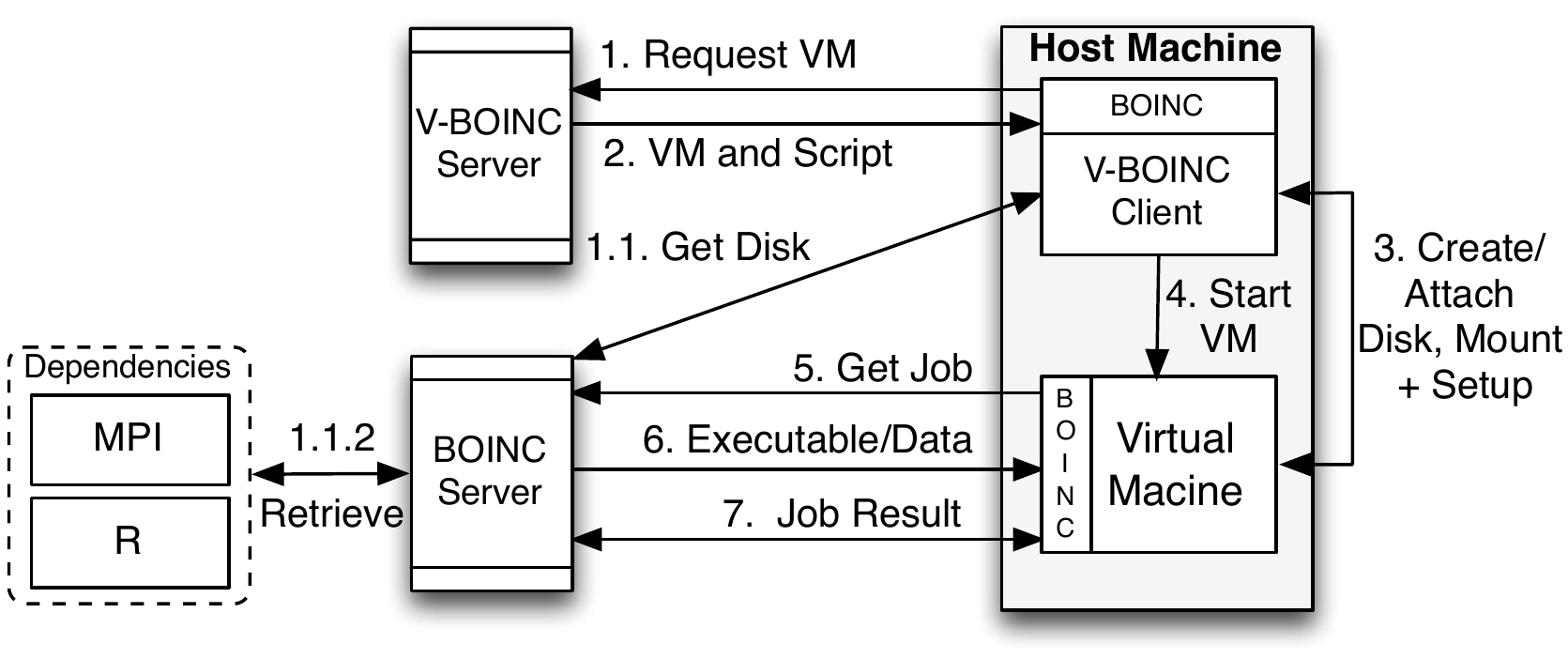}
  \end{center}
 \caption{V-BOINC Implementation Overview}
\end{figure}

Upon a volunteer user submitting the details of the BOINC scientific project they wish to attach to via the V-BOINC Client (e.g the project server URL and their BOINC project weak account key), the host BOINC client is instructed to request a virtual machine image (1). Concurrently, the V-BOINC Client probes the BOINC server to determine if any dependencies exist for the specified project (1.1). If so, a VDI (or \textit{.vdi}) file containing the dependencies is transferred to the V-BOINC Client via \textit{curl}; we assume that developers of BOINC projects who wish to deploy applications with dependencies are prepared to create a VDI file containing the dependencies and make this publicly available on the BOINC Server to allow the V-BOINC Client to determine whether a DepDisk needs to be downloaded. 

Concurrently while a DepDisk is downloading, the virtual machine image and an executable script are downloaded to the host BOINC client (2); both download processes must complete before proceeding to the next step. The V-BOINC Client either attaches the DepDisk, if the application is found to have dependencies, or alternatively creates an empty disk and mounts this to the virtual machine image (3). The virtual machine image is then started (4) allowing it to request (5) and receive (6) BOINC jobs and return job results (7).

\subsection{Lightweight, Flexible and Robust VMs}
The purpose of attaching/creating mountable DepDisks above (1.1.2/3) is well justified for a number of reasons. As opposed to relying on volunteer host dependencies where packages must be present and in a specific location on a volunteer machine; in turn limiting the number of hosts available to a specific project due to the many different host configurations possible; the use of mountable disks makes it an easy and effective method for applications with dependencies to run. Without the use of virtualization, software packages (e.g MPI, R, Java etc) could be transferred and utilized via regular BOINC, however one could not take advantage of virtualization.

To reduce the bandwidth consumed by transferring V-BOINC virtual machine images to volunteer hosts, the virtual machines has been stripped of all unnecessary components such as Linux swap space and unneeded packages. As a consequence, no extra disk space exists hence why mountable disks are required not only for adding application dependencies but for adding disk space for applications to consume. Where no dependencies are required, a fresh disk is locally created on the volunteer host and mounted. In both cases, Linux swap space is replaced to ensure the performance of the virtual machine is not degraded. As a result of distributing stripped virtual machines, no bandwidth is wasted by transferring these images with unused disk space.

To create the smallest usable virtual machine image possible, we use the VirtualBox Fixed Disk Image (FDI) type as opposed to the Dynamic Disk Image (DDI). The former is of fixed size and the latter has the capability to grow according to how much is stored upon it, up to a specified maximum; this image however does not decrease in size when items are removed from a virtual machine. Our virtual machine VDI uses the FDI image type for one important reason: the size of the DDI image is difficult to control and keep as small as possible. Also, for example, an FDI file with the OS components installed could size at 681MB however with the same components installed, a DDI could be 700MB. It is important to keep the virtual machine VDI at an absolute minimum to reduce the data transferred and stored on the host. The current size of our virtual machine VDI is 649 MB uncompressed and 207 MB compressed.

On the other hand, DepDisks use the DDI type to minimize the initial storage required on the host. For example, when the virtual machine image is downloaded and the DepDisk attached, the minimal storage possible is consumed due to the combination of different disk types used. By essentially partitioning a virtual machine over two VDI files, we ensure that when a user attaches to another BOINC project, a new DepDisk need only be `plugged' in to the virtual machine as opposed to downloading both a new virtual machine image and DepDisk.

\subsection{Taking Control}
After the virtual machine image has been transferred to the volunteer machine via the host BOINC client --- an operation that would only take 3 minutes assuming that the current average UK bandwidth of 9Mbps \cite{ofcom} applies --- it must be unpacked, configured and started; a process which is performed both by the instantiation script downloaded in step (2) of Figure 1 and the V-BOINC Client. The instantiation script simply: 

\begin{itemize}
\item Decompresses the virtual machine image tar file. 
\item Signals the V-BOINC Client to take control of the instantiation process.
\end{itemize}

When signalled by the instantiation script, the modified BOINC client as part of the V-BOINC Client:
\begin{itemize}
\item Registers the virtual machine image with VirtualBox.
\item Creates a fresh VDI or attaches a pre-created DepDisk to the virtual machine.
\item Starts the virtual machine image.
\item Takes periodic snapshots once the virtual machine is running.
\item Waits for the virtual machine process to finish. This firstly shows to the user that the virtual machine process is still running if they use the BOINC Manager and that any virtual machine errors are caught during execution which can then be uploaded to the server afterwards for debugging.
\end{itemize}

Once the virtual machine process is running, further complexities are introduced as a second BOINC client located on the virtual machine needs to be controlled from the host to execute typical BOINC commands such as requesting tasks and uploading results; this is performed by using the \textit{boinccmd} command line tool through the V-BOINC Client GUI. Figure 2 shows how the V-BOINC Client GUI must interact with both BOINC clients and the VirtualBox API.

\begin{figure}[h!]
  \begin{center}
\includegraphics[width=0.45\textwidth]{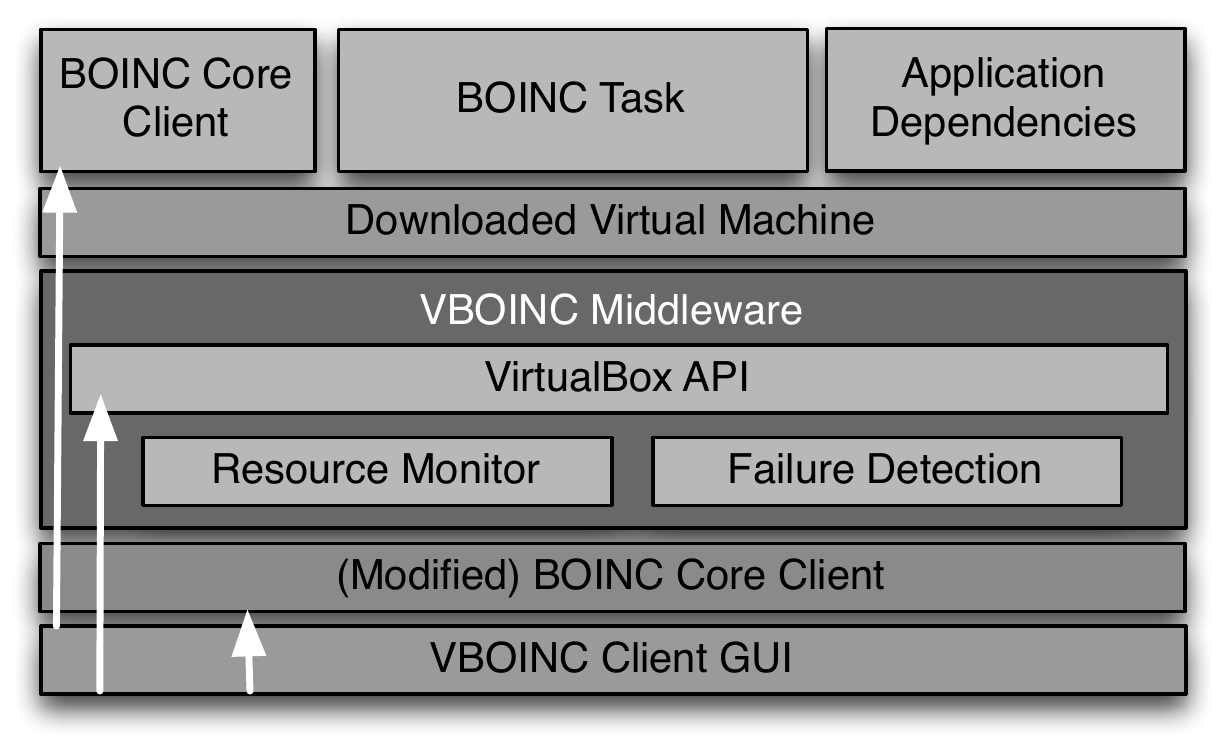}
  \end{center}
 \caption{V-BOINC Volunteer Host Components}
\end{figure}

The V-BOINC Client GUI provides a similar interface to that of the BOINC Manager, offering options to control either BOINC client's state to either running, suspended and halted via the \textit{boinccmd} component. For example, if a user wishes to suspend a job running on the virtual machine, one has to use the \textit{suspend} directive via the \textit{boinccmd} tool on the virtual guest. BOINC offers other command options such as: \textit{reset}, \textit{detach}, \textit{update}, \textit{resume}, \textit{nomorework} and \textit{allowmorework}. These commands must be passed to the V-BOINC Middleware component which wraps them in a VirtualBox API method call to the \textit{guestcontrol} function and executes them on the virtual machine; the virtual machine has Guest Additions installed to allow this.

These commands will control a BOINC job's execution within the virtual machine process however controlling the virtual machine itself is more complex as the host BOINC client cannot (easily) control separate non-BOINC processes. For example, the above \textit{boinccmd suspend} command would not suspend the virtual machine process if executed locally on the host. Commands such as these must be performed via the VirtualBox API by calling the \textit{controlvm} component. Additionally, the Middleware component also provides resource monitoring and virtual machine failure detection to inform the user at real time, the current state of V-BOINC. 

\subsection{Checkpointing and Recovery}
To ensure the continuity of BOINC applications, the modified BOINC client implements periodic virtual machine checkpointing and recovery, with the interval between snapshots chosen by the volunteer user. In the case of any errors occurring on the volunteer host, or the host simply terminates, the latest snapshot can be recovered and the project developer can be reassured that the computation will complete without the need for implementing application checkpointing. VirtualBox makes checkpointing simple by calling the \textit{snapshot} component of the VirtualBox API. Executing this command places the appropriate snapshot files in the \textit{Snapshots} folder where the virtual machine image is located. The files created when checkpointing via VirtualBox are: 

\begin{itemize}
\item A copy of the virtual machine settings. These settings include the hardware configuration such as the memory allocated to the machine as well as any attached disks.

\item The current state of all VDI's attached to the virtual machine. VirtualBox implements this by storing differencing images; images which store all write operations after a snapshot is taken.

\item The current state of memory if a snapshot is taken while the virtual machine is running. This memory state file can be quite large --- up to the memory size allocated to the machine --- and is dependent on the application memory usage. Allocating less memory, limits the size of the memory dump file but reduces application performance for those dependent on memory.
\end{itemize}

To restore a snapshot, the correct differencing image is activated and the current snapshot/virtual machine state is deactivated. To reduce the storage space consumed on the host, previous stale snapshots files that are not required are deleted by V-BOINC.  

\section{Experiments and Results}
We now outline the experiments performed to firstly show the achievable resource performance of V-BOINC when compared to regular BOINC and secondly to outline our use case showing that the V-BOINC framework can be used for computations requiring dependencies. Thirdly we show what effect of periodic system-level checkpointing has on the valuable storage space reserved for BOINC jobs and on the BOINC job itself.  All experiments were performed on an OptiPlex 790 host with two Intel i3-2100 Core 3.10 GHz processors and 3.8 GB of memory. By default, the V-BOINC virtual machine is set to use the hardware assisted virtualization instruction sets VT-x/AMD-V and the default values of using 1 GB of RAM and 1 processor are increased to the maximum VirtualBox allows.

\subsection{BOINC vs V-BOINC}
To evaluate V-BOINC, we measured the performance of V-BOINC when compared to regular BOINC. This was performed by running a series of benchmarks and a use case application and collecting their execution times.

\subsubsection{Benchmark Performance}
We used six benchmarks, shown in Figure 3, each with different resource usage demands to demonstrate the performance of a range of workloads. Each benchmark was executed ten times and the average of these figures was plotted. We display 95\% confidence intervals to show that in most cases, the true mean will lie within the specified range. 

\begin{figure}[h!]
  \begin{center}
\includegraphics[width=0.47\textwidth]{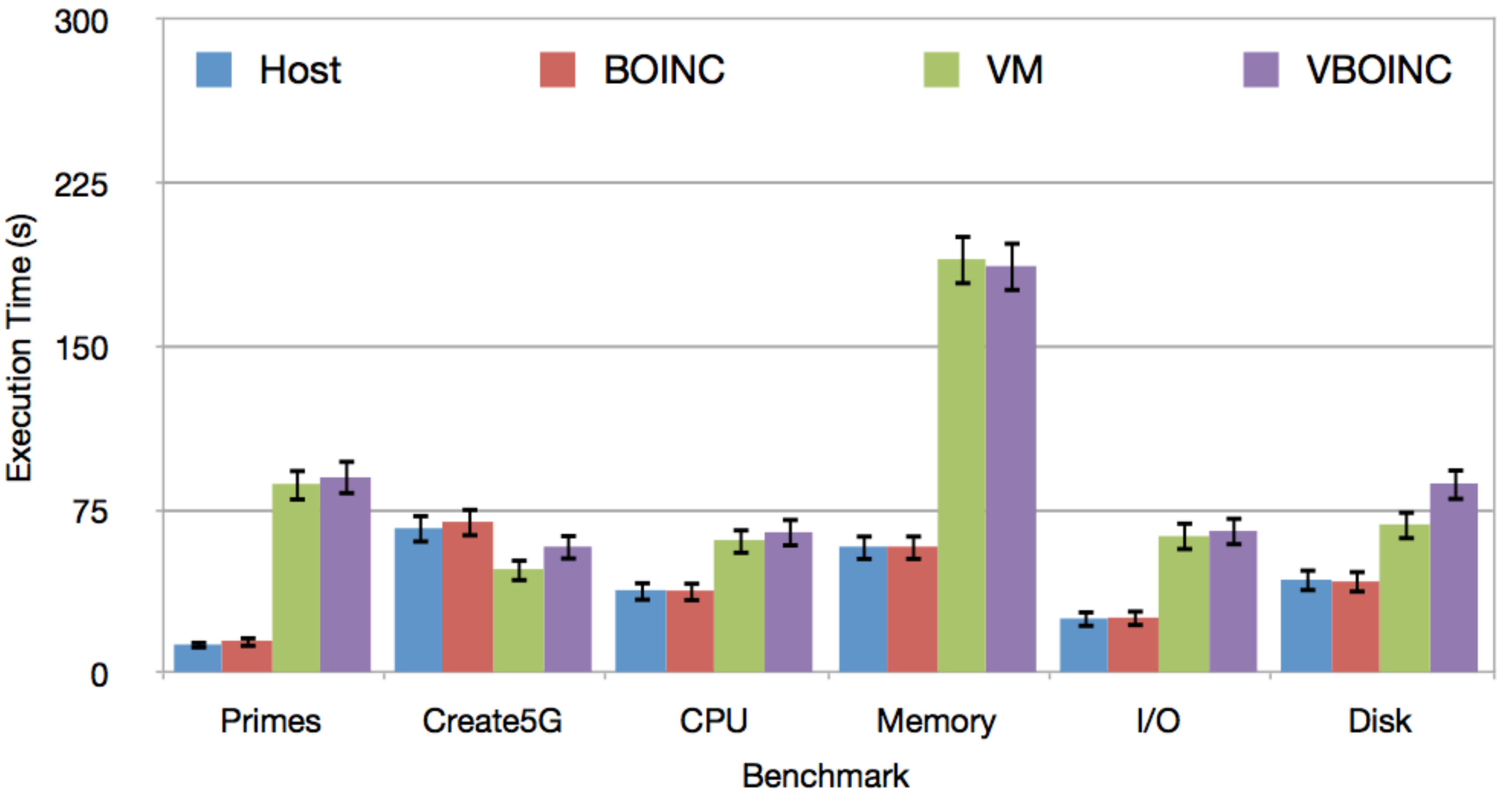}
  \end{center}
 \caption{V-BOINC Benchmark Execution Times}
\end{figure}
\textbf{Primes} is a CPU intensive application used to calculate the first 300 prime numbers. \textbf{Create5GB} is a memory and I/O intensive function used to create a file of 5 GB using the Linux function \textit{dd}. \textbf{CPU, Memory, I/O} and \textbf{Disk} are modified versions of the Stress workload generator \cite{stress} to strain each of the resources up to a specified number of iterations. Each benchmark is then run over four different platform configurations: 
\begin{enumerate}
\item execution just on the \textit{Host} without the use of BOINC.
\item execution on the \textit{Host} using BOINC.
\item execution just on the V-BOINC virtual machine without the use of V-BOINC.
\item execution on the V-BOINC virtual machine using V-BOINC.
\end{enumerate}

Figure 3 shows the executions times obtained by running the benchmarks in each case above. Firstly we see that the overhead of BOINC is negligible when comparing cases (1) and (2). Secondly and most importantly, we see that in most cases V-BOINC is slower than traditional BOINC with the exception of the Create5GB benchmark. Thirdly we see that the implementation of V-BOINC introduces little overhead when comparing cases (3) and (4). This shows that the performance difference between BOINC and V-BOINC is introduced by virtualization alone and not the implementation. 

This slowdown is caused by many factors relating to the virtual machine settings and hypervisor. When a virtual machine image is registered with VirtualBox, one must specify the memory, number of CPU's to use as well as a CPU execution cap, i.e only use 90\% of the processor for example. However because the virtual machine is not able to use the full amount of memory and processing power available to the host machine, it is predictable that V-BOINC would perform slower; only 2.9 GB of RAM could be allocated to the virtual machine, hence explaining why memory intensive benchmarks perform much slower. 

Our memory benchmark execution time above uses 2.5 GB of memory; approximately 66.9\% and 85.2\% of the total available host and virtual machine memory respectively. If we normalize the percentage of memory used to 66.9\% for each host, the execution time difference reduces from 190 seconds to approximately 160 seconds, showing the true virtualization overhead. However, the remaining memory intensive benchmark Create5GB shows that not all applications may run slower when using virtualization and this is dependent on the internal components of the application. We can only assume that the hypervisor's caching strategy is better than that of the underlying system.

Similar to the memory deficit, the processing power available to the virtual machine is lower than the total available to the underlying host. This is caused by the resources used by processes running and supporting the hypervisor on top of those running the Operating System. Hence the performance differences between host and virtual machine executions can be partly attributed to the maximum settings VirtualBox allows for any particular virtual machine but also the performance of VirtualBox itself where others have found the performance difference much slower than execution upon the host \cite{Ferreira2011}\cite{Younge2011}\cite{Domingues2009}.
\begin{table*}[ht]
\begin{center}
\caption{Snapshot Files and Times per Benchmark}

\begin{tabular}{|l|c| c|c|c|}
	\hline
$\textbf{Benchmark}$ &  $\textbf{Snapshot Time (s)}$   & $\textbf{Memory Size (MB) }$ & $\textbf{DepDisk Snapshot Size (KB) }$  & $\textbf{VM Snapshot Size (KB)}$\\ \hline
CPU & 1.1779 &	86.9 & 36 &	8  \\ \hline
Memory & 1.7142 & 56.76 & 36 & 8 \\ \hline
I/O & 0.9425 & 43.57  & 36 & 8\\ \hline
Disk & 24.6023 & 1126.4 & 54374.4  & 8 \\ \hline
Primes & 1.2153 & 98.1 & 36  & 8\\ \hline
SPRINT & 31.4665 & 2334.72  & 36  & 8\\ \hline
\end{tabular}
\end{center}
\end{table*}

\subsubsection{Case Study: SPRINT-R}
To illustrate that V-BOINC can not only execute standalone applications and to also show the performance achieved of a real use case application, we execute the Simple Parallel R INTerface (SPRINT) \cite{Piotrowski2013}, which has MPI and the statistical package R as dependencies, on V-BOINC. SPRINT is a package providing parallel functions of R allowing data to be analysed over multiple processors rather than performing the computation on a single node and was chosen as it already is in wide use on computing clusters. The support and analysis of real users running such applications on the V-BOINC infrastructure will be undertaken in the near future however.

For our experiment, we used SPRINT's \textit{pcor} which is the parallel version of the R serial function \textit{cor}. As its name may suggest, it performs correlation on a given data set; our data set is randomly generated with 11000 genes (rows) and 321 samples (columns). First this data must be loaded (Load) into R and then executed (Exec). Figure 4 shows the execution times of these operations respectively when two SPRINT processes are spawned. Again we provide a comparison between running the application via a variety of configurations, i.e Host, BOINC, VM and V-BOINC.
\begin{figure}[h!]
  \begin{center}
\includegraphics[width=0.45\textwidth]{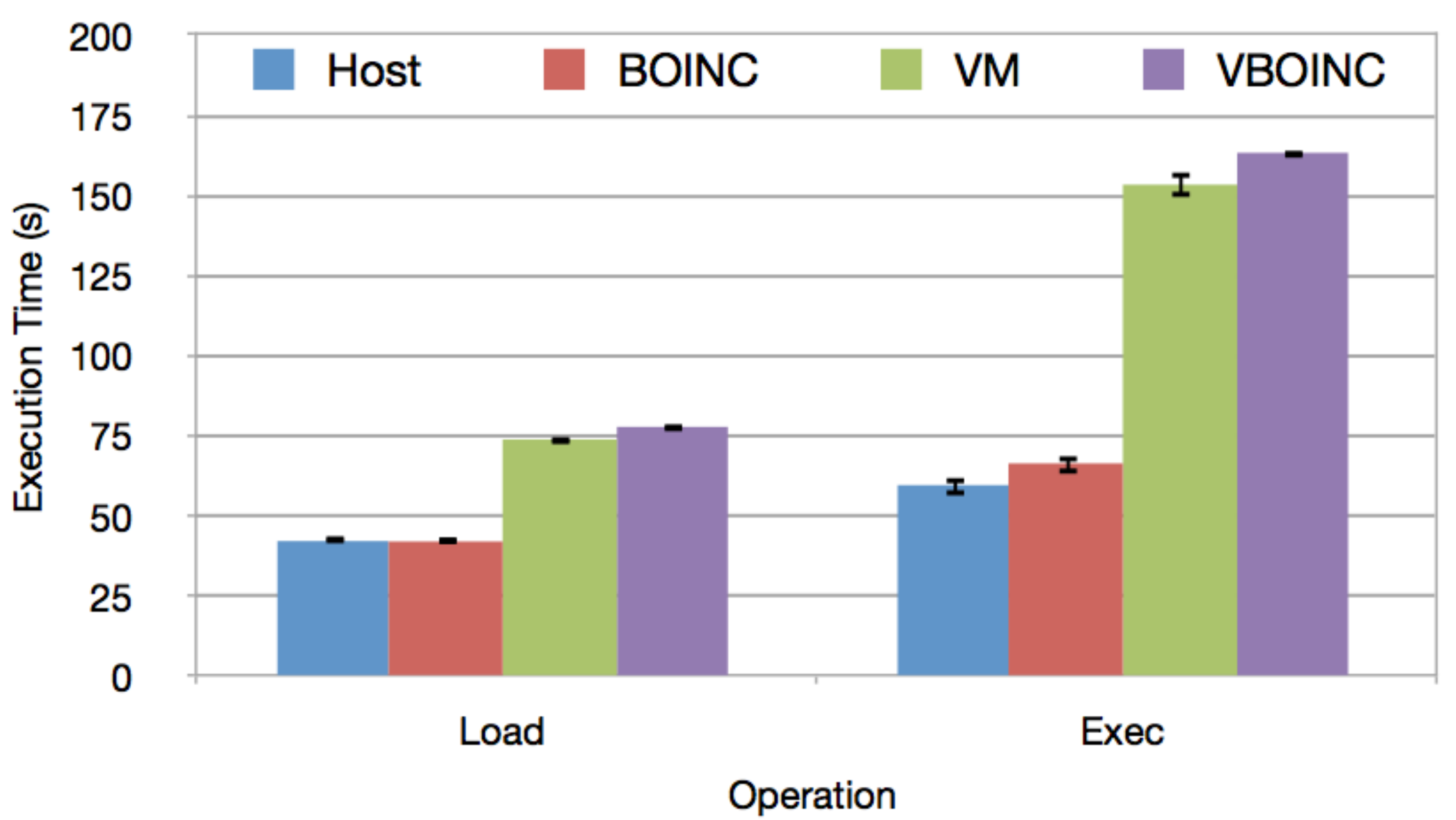}
  \end{center}
 \caption{SPRINT Data Load and Execution Times}
\end{figure}

Figure 4 depicts three results similar to those outlined in Figure 3. Firstly that running SPRINT on BOINC shows little or no overhead when compared to running SPRINT on the host itself. Secondly, the overhead of the V-BOINC implementation is also minimal where little difference can be seen when running SPRINT via the cases (3) and (4) above. Thirdly and most importantly, we see the performance difference between V-BOINC and regular BOINC, with the overhead of virtualization causing the time for loading and execution to approximately double and triple respectively in these cases; a fact that must be accepted when using virtualization to solve other problems.

\subsection{The Effect of Checkpointing}
To enable BOINC project developers to omit application-level checkpointing from their code, V-BOINC provides periodic checkpoints. However, because the storage space BOINC is permitted to use is potentially limited by the volunteer specified preferences, this makes it extremely valuable. To determine the likely storage space consumed by our system-level checkpointing approach, we executed a series of benchmarks representing different workloads while taking per one minute checkpoints over a ten minute period and recording the storage space consumed. Table 2 shows the average values obtained for the time taken to perform a snapshot, the size of memory dump file and the actual snapshot VDI sizes of the DepDisk and virtual machine. The runs were performed using the V-BOINC virtual machine with an attached 8 GB DDI DepDisk containing experiment files and the necessary dependencies for SPRINT. 

Firstly, we see that in four of the six resource intensive benchmarks (CPU, Memory, I/O and Primes), the average snapshot time is minimal at approximately 1 second. In these cases, we also see that the memory dump file size is lower than 100 MB and that the snapshot VDI sizes of the DepDisk and virtual machine also remain small at 36 KB and 8 KB respectively; the lowest possible snapshot sizes for these two disks. This shows that the DepDisk or VM VDIs are not written to during execution, where only CPU, memory and I/O resources were used. The remaining Disk and SPRINT benchmarks show different results where snapshot times and memory dump file sizes are larger. This is caused by a large amount of memory consumed in both cases and a large amount of writes to disk in the former.  In these cases, the largest memory dump file recorded was 2.28 GB using SPRINT and 1.1 GB using the Disk-intensive benchmark; this benchmark also has the largest DepDisk snapshot VDI size of approximately 53 MB. 

These results show that applications that do not write to disk or perform lots of memory operations (e.g cache writes etc) are unlikely to consume large amounts of storage space on the volunteer host when periodic snapshots are taken. However applications which intensively perform memory or disk operations are likely to produce larger memory dump and snapshot files. This is reassuring as typically BOINC applications tend to be CPU intensive operating over little data (e.g SETI@Home uses about 10MB per host \cite{Anderson2006}) hence the checkpointing process should be quick and consume very little storage space.

\subsection{A Note on Server Performance}
Similar to the performance degradation caused by virtualization on the volunteer host, we expect the performance of the V-BOINC server to be less than that of a regular BOINC project server deployed on the same host. Previous research shows that a BOINC server hosted on a single inexpensive computer can distribute up to 8.8 million tasks per day with the CPU and network bandwidth being the main bottlenecks \cite{Anderson2005}. In the case of V-BOINC, we expect that the number of tasks per day the server can distribute will be significantly lower than that of a regular BOINC server, with the network bandwidth being the major bottleneck when volunteer BOINC clients request a virtual machine image to be downloaded on their machine.

To alleviate this problem, BOINC server administrators currently solve CPU and network bottlenecks by replicating a server across a larger number of machines. Dependent on the popularity of V-BOINC, we may employ this replication mechanism over multiple Amazon EC2 regions to reduce the distance between volunteer user and a V-BOINC server. Furthermore, BOINC clients are designed to employ exponential back off of requests to the server, hence the V-BOINC server should rarely receive a large number of requests that cause it to experience failures due to CPU and network bandwidth bottlenecks.

\section{Conclusions and Future Work}
V-BOINC is a tool providing solutions to the drawbacks of regular BOINC by allowing project developers to port their application only to the V-BOINC virtual machine and omit application-level checkpointing from their code. Developers with applications that have dependencies can easily utilize V-BOINC where users of regular BOINC cannot (easily) run such applications. Finally end user worries relating to security and untrustworthy applications are also solved via the sandbox nature of virtual machines. Note that V-BOINC does not currently deal with providing correct credit to BOINC users nor does it accurately adhere to user based preferences; these features as well as advanced resource monitoring and failure detection will be present in future versions of V-BOINC. 

One will find that the design and implementation of  V-BOINC plays a major role on how regular BOINC applications and those with dependencies can easily be run upon V-BOINC. In the former case, our inner virtual machine BOINC client allows regular BOINC applications to be run in the virtual presence without modification and furthermore, four stage transfers between the virtual machine and host do not occur as implemented in other research \cite{Ferreira2011} \cite{Marosi2010}. In the latter case, the attachable disk mechanism allows dependencies to be mounted automatically and snapshots of this disk can be taken and restored upon virtual machine termination or failures.

We have also shown how the performance of V-BOINC compares to regular BOINC and how the implementation of V-BOINC introduces a negligible overhead. As expected the performance of regular BOINC is better than that of V-BOINC's due to the virtual machine overhead, however the actual overhead caused by the implementation of V-BOINC is negligible when compared to running the same application on a standalone virtual machine. However one must weigh up the advantages of V-BOINC compared to the increased performance of traditional BOINC and whether the performance cost from virtualization is acceptable for volunteer computing. Investigating this with real volunteer users, application communities and different hypervisors such as QEMU/KVM and VWWare Player is worthy of extensive future investigation.

V-BOINC will continue to be optimized and developed to include the omitted features mentioned and is motivated by the fact that many users have downloaded the framework. For those who wish to test V-BOINC, our server and client components as well as links to our project documentation and online Amazon EC2 service can be found at \cite{vboinc_info}.

\bibliographystyle{IEEEtran}
\bibliography{Papers_Read,Other_Refs,My_Papers}

\end{document}